\documentclass[manuscript,screen]{acmart}
\usepackage{enumitem}
\setlist[itemize]{leftmargin=*}
\authorsaddresses{}
\usepackage{tabularx}
\setcopyright{none}             % Clear ACM copyright text
\renewcommand\footnotetextcopyrightpermission[1]{} % Removes footnote block
\AtBeginDocument{%
  }

\setcopyright{acmlicensed}
\copyrightyear{2026}
\acmYear{2026}
\acmDOI{XXXXXXX.XXXXXXX}
\acmConference[]{}{June 03--05,
  2018}{Woodstock, NY}
\acmISBN{978-1-4503-XXXX-X/2018/06}

\begin{document}


% \title{What Can We Become? Expanding Ontological Imaginations in Personal Sensing}
% \title{Design for Ontological Boundary Negotiation: An Exploratory Study with Two Open-ended Personal Sensing Probes}
\title{Negotiating Ontological Boundaries in User-Authored Personal Sensing Systems}
% \title{Drawing Boundaries, Making Selves: An Exploratory Study of Living Classifications in User-Authored Personal Sensing}


\author{Nava Haghighi}
\email{nava@cs.stanford.edu}
\affiliation{%
  \institution{Apple Inc. \& Stanford University}
  \streetaddress{450 Serra Mall}
  \city{Stanford}
  \state{CA}
  \country{USA}
  \postcode{94305}
}

\author{Danielle Olson}
\affiliation{%
 \institution{Apple Inc. }
 \city{Cupertino}
 \state{CA}
 \country{USA}}
\email{danielleolson@apple.com}

\author{Halden Lin}
\affiliation{%
 \institution{Apple Inc. }
 \city{Cupertino}
 \state{CA}
 \country{USA}}
\email{halden@apple.com}

\author{Erdrin Azemi}
\affiliation{%
 \institution{Apple Inc.}
 \city{Cupertino}
 \state{CA}
 \country{USA}}
\email{erdrin@apple.com}

\author{Gierad Laput}
\affiliation{%
 \institution{Apple Inc.}
 \city{Cupertino}
 \state{CA}
 \country{USA}}
\email{gierad@apple.com}

\author{Kayur Patel}
\affiliation{%
 \institution{Apple Inc.}
 \city{Cupertino}
 \state{CA}
 \country{USA}}
\email{kdpatel@gmail.com}

\author{James Landay}
\affiliation{%
 \institution{Stanford University}
 \city{Stanford}
 \state{CA}
 \country{USA}}
\email{landay@stanford.edu}

\renewcommand{\shortauthors}{Haghighi, et al.}

\begin{abstract}
Designed artifacts are ontological, shaping, and at times limiting, what becomes possible or imaginable. One path toward mitigating such foreclosures is giving people power over how systems are designed and built. Despite decades of scholarship around systems that enable such authorship, these systems are often evaluated on whether or not they are usable, useful, or technically feasible, leaving questions of ontological boundary negotiation, unexamined. 
We design two open-ended probes that utilize a Wizard of Oz technique to enable the experience of training a personalized machine learning system on phenomena people define themselves. 
In a week-long exploratory study, participants use one of two probes in the course of their everyday lives.
We identify four sites where ontological boundaries were negotiated; the boundaries of a phenomena, the subject as part of relations, what is signal and what is noise, and the objectivity of data. 
We offer starting points for supporting boundary negotiation through design and discuss open-ended probes as a method for ontological design.

\end{abstract}

\begin{CCSXML}
<ccs2012>
   <concept>
       <concept_id>10003120.10003121.10003126</concept_id>
       <concept_desc>Human-centered computing~HCI theory, concepts and models</concept_desc>
       <concept_significance>500</concept_significance>
       </concept>
   <concept>
       <concept_id>10003120.10003121.10003122.10003334</concept_id>
       <concept_desc>Human-centered computing~User studies</concept_desc>
       <concept_significance>300</concept_significance>
       </concept>
 </ccs2012>
\end{CCSXML}

\ccsdesc[500]{Human-centered computing~HCI theory, concepts and models}
\ccsdesc[300]{Human-centered computing~User studies}

\keywords{ontologies, ontological design, diffraction, boundary negotiation, personal sensing, personal informatics, probes}

%   \includegraphics[width=\textwidth]{Sections/Assets/manyworlds.png}
%   \caption{Negotiating the .}
%   % \caption{Each ontological dimension of the design space of personal informatics can connect us to new possibilities for relating to our selves in expansive ways, opening portals to the world of many worlds we already live in. Rather than a prescriptive framework, we define a provisional generative framework with four dimensions (define, detect, represent, engage), share the ontological questions posed by each dimension, and the possibilities that get opened up by those questions.}
%   \Description{six marble-like spheres metaphorically representing the worlds opened up by different ontological possibilities. The spheres are connected to each other via four different line types and colors. Each line type represents one of define, detect, represent, or engage. }

% \received{1 February 2026}

\maketitle

\section{Introduction}
Design is \textit{ontological}~\cite{winograd1986understanding, willis2006ontological}, shaped by the assumptions that bound what people imagine, articulate, or deem possible~\cite{haghighi2025ontologies}. 
Thus, designed artifacts do not neutrally serve human needs, but are sociotechnical artifacts that actively participate in shaping worlds, making some worlds possible and imaginable, while foreclosing others~\cite{costanza2020design, willis2006ontological, winograd1986understanding, fry2009design, winner2017artifacts}. 

One path toward mitigating such foreclosures is giving people power over how systems are designed and built~\cite{power_li}, enabling them to author their own systems and/or shape them after deployment~\cite{ehn2008participation, masnick2019protocols}.\footnote{In this work, I will refer to the category of systems that are authored by the user and/or shaped by them after deployment as user-authored.}
Recent advances in Generative AI are making user-authorship increasingly possible for a broader group of people in various domains (for example see~\cite{litt2025malleable}), offering a unique opportunity for surfacing and questioning the ontological boundaries that fixed systems and platforms naturalize. 
However, user-authorship is not guaranteed to expand existing ontological boundaries. Whether user participation in authoring systems can actually negotiate and expand the assumptions embedded in designed systems through authorship, under what conditions, and with what limitations, remains an open question. Thus, in this paper we ask: \textbf{how can design create and support the conditions for ontological boundary negotiation in user-authored systems?}

Personal sensing is a particularly productive site for examining this question. Personal sensing systems draw boundaries around phenomena to be able to work with them, for example deciding what counts as stress, focus, or fatigue, what counts as signal and what is noise, and even, what a person \textit{is}~\cite{leahu2014freaky, 10.1145/3772318.3791814}. A longstanding body of critical work has documented the boundaries that often get taken for-granted in these systems. For example, these systems often center the human as a bounded individual with objectively measurable properties who is capable of self-improvement through data-driven behavior change~\cite{lupton2016quantified, knowthyself, 10.1145/3772318.3791814, elsden2016fitter}. 
HCI and personal sensing researchers have made meaningful steps toward giving users agency to author their own sensing systems and/or algorithms~\cite{omnitrack, dey2004cappella, hartmann2007authoring, simard2017machine, amershi2014power}.
But these systems are often evaluated on whether or not they are usable, learnable, or useful, leaving the question of whether and how user-authored systems can enable ontological boundary negotiation, unexamined. To address this gap, we examine the \textbf{negotiation of ontological boundaries in user-authored personal sensing systems}.

We build two open-ended probes that enable the experience of training a personalized machine learning system on phenomena people define themselves. 
Because open-ended systems that allow real-time, on-device training on user-defined phenomena do not yet exist, we use a Wizard of Oz technique to decouple the authoring experience from present-day technical constraints. To this end, participants are told that the data collected during the study will be used to train a model \textit{after} the study, rather than in real-time and on-device, while offering a UI that approximates the on-device training process through visual feedback on training examples collected. The first probe, \textbf{EventMarker}, simulates training an algorithm to detect marked experiences or events through sensor data, connecting lived experience to what a sensor can capture. The second, \textbf{PatternFinder}, simulates training an algorithm to detect patterns of interest observed and labeled by the user in their physiological data, connecting sensor data to lived experience. 
The two probes differ in the level of abstraction at which authoring begins so as to examine how that shapes which boundaries become visible and negotiable.

In a week-long exploratory study, eight participants use one of two probes in the course of their everyday lives, defining their own categories, identifying their own phenomena of interest, and imagining — outside of the constraints of what they deem technically feasible — how they might use such a system. 
To analyze the results, we attend to moments where boundaries that personal sensing systems typically fix get negotiated by participants in the course of using the probes, and what factors support or foreclose such negotiations. 
We identify four sites of ontological boundary negotiation: (1) negotiating the boundaries of a phenomena, (2) negotiating the subject as part of relations, (3) negotiating what is signal and what is considered noise, and (4) negotiating the objectivity of data. 
Reflecting on the contexts in which a boundary was negotiated or was moved toward fixity, we offer starting points for supporting boundary negotiation through design, discuss the politics of designing for provisionality, and open-ended probes as a method for ontological design.

This paper examines related work around ontologies, participation in design and user-authored systems, and personal sensing, describes the two probes, presents findings from a week-long study with eight participants, and concludes by discussing four sites for boundary-negotiation and the implications for ontological designing of personal sensing systems.

\section{Related Work}
% PI in general, personalized ML in PI, critical perspectives in PI, and a section on ontologies in design

In this section, we offer a brief introduction to ontologies in design, and share an overview of participation in design and self-authored systems, with a focus on the domain of personal sensing.

\subsection{Ontologies and Ontological Design}
As a branch of philosophy, ontology refers to the nature of ``being''~\cite{malik2021critical}. 
Within AI and computing, Gruber borrowed the term ontology from the philosophical definition to introduce systems for formally representing knowledge~\cite{gruber1991role, gruber1995toward}.
In this work, we use the term ontologies, in the plural, as used by critical humanities scholars and summarized by~\citet{haghighi2025ontologies} as ``what we allow ourselves to talk or think about'' or imagine possible. 
Ontology, in this use of the word, was introduced in HCI by~\citet{winograd1986understanding}, arguing that the rationalistic traditions in computer science do not take into account ontologies. Later expanded by~\citet{willis2006ontological}, ontologies have influenced HCI research and theory. For example, in feminist design, Bardzell and Bardzell share how the design of the ``Hoozier'' cabinet further identified women with household work~\cite{bardzell_feminist_2010}. 
More recently,~\citet{haghighi2025ontologies} argue that critique of AI systems has concentrated on values and bias while neglecting ontological commitments underlying AI systems. 

In this work, we examine how and if the increased agency afforded to people through user-authored systems can result in moving beyond the ontological limitations of existing systems, and what role design might continue to play in that context. 

\section{Participation in Design and User-authored Systems}

Across HCI, CSCW, and participatory design, prior work has examined how people adapt technologies, sometimes beyond their intended design~\cite{power_li, li2025reimagining}, and how systems and infrastructures can be built to support such adaptations. This work includes systems that users can extend after deployment through meta-design~\cite{fischer2004meta}, documentation of how users complete design in use through appropriation theory~\cite{dourish2003appropriation, dix2007designing, carroll2004completing}, and technical affordances for doing so as discussed in end-user development~\cite{cypher1993watch, lieberman2006end, lieberman2001your, ko2011state}. 
Design scholars have also made the case for open, ambiguous artifacts that invite, and in a way demand, interpretation~\cite{gaver2003ambiguity, sengers2006interpretation, sanches2019ambiguity}. 
Furthermore, participatory design has brought attention to infrastructuring in design~\cite{karasti_infrastructuring_2014, ehn2008participation}, discussing the role of the designer not simply as designing an artifact but building infrastructures that sustain participation over time. 
The recent advances in generative AI are taking the possibilities for such participation in design a step further, enabling visions such as malleable software in practice~\cite{litt2025malleable}. 
Despite this growing interest in participation in design and user-authored systems, what scaffolding can support an expansion of \textit{ontological possibilities} in such systems has not been examined. 

Within this work, one domain that is particularly productive for examining the ontological assumptions that get taken for granted is personal sensing. 
For decades, the assumptions underlying these systems have been examined and critiqued. Specifically, personal sensing systems often assume the human to be a bounded individual that has properties that can be objectively measured and with the right type of intervention, controlled~\cite{fogg, knowthyself, lupton2016quantified, elsden2016fitter, 10.1145/3772318.3791814}. 
In response, reflective design~\cite{reflective} calls for critical examination of the values embedded in technology, with work such as ``Fit4life''~\cite{purpura2011fit4life} examining what happens when this view of personal sensing gets taken to its extreme. 
However, despite the rich body of critical work (for example, see~\cite{dourish2001action, somadesign, postphenomenology_PI, homewood, 10.1145/3772318.3791814}), an analysis of over 100 commercial apps found limited support for deeper, transformative reflection~\cite{reflection, fleck2010reflecting}.
More recently,~\citet{10.1145/3706598.3713940} examine three case studies of present-day well-being technologies and find these limitations further reinforced by machine learning models underlying these technologies. 

Given the relatively limited commercial impact of the critiques of personal sensing systems, increased user authorship and agency can enable people to move beyond the limitations imposed by the current fixed systems~\cite{reflection, 10.1145/3706598.3713940}. 
Within personal sensing, researchers have explored what openness and user authorship might look like in practice. Dey et al.'s CAPpella~\cite{dey2004cappella} and Hartmann et al.'s Exemplar~\cite{hartmann2007authoring} are examples of early systems that give people the ability to define their own categories and train their own systems of classification. OmniTrack~\cite{omnitrack} extended this by enabling users to compose their own trackers, combining manual and automated tracking. 
Sanches et al.~\cite{sanches2019ambiguity} showed that deliberately ambiguous physiological data can give rise to multiple proto-practices organized around participants' own interpretations.
Thus, in this work, we take interest in examining how design can make room for a negotiation and expansion of ontological boundaries in and through personal sensing systems that users can continue to participate in and shape over time.

\section{Two Open-ended Probes for Exploring Personalized AI-based Sensing Systems} \label{design_probes}

To examine how design can create and support the conditions for ontological boundary negotiation in personal sensing systems, we designed two open-ended probes~\cite{boehner2012probes, hutchinson2003technology, gaver_design_1999}. 
Open-ended probes give access to what people are thinking and doing in absence of prescribed categories or fixed tasks, revealing in part, the boundaries people are operating within. 
Furthermore, using probes for an extended period offers access to the temporal evolution of people's engagement with the probes and places where existing boundaries were negotiated. 

This approach differs from existing speculative or critical sensing projects in that we did not intend for the probes to offer a critical alternative or provocative approach to, or imagination of, existing sensing systems (for a great example of this approach, see Fit4life~\cite{purpura2011fit4life}). Instead, the two probes resemble the dominant approaches already existing in personal sensing, both in research and commercial systems~\cite{reflection,epstein2020mapping, dey2004cappella, hartmann2007authoring}.
Namely, the probes are distinguished by the level of information abstraction at which participants engage with their data. Personal sensing systems mediate between different levels of information abstraction, for example connecting low-level physiological signals such as heart rate, electrodermal activity, or accelerometer data, with high-level mental constructs such as stress, fatigue, or focus. 

The probes offer the scaffolding for participants to engage with their data either at higher or lower levels of information abstraction. 
\textbf{EventMarker}, not unlike an early system like a CAPpella~\cite{dey2004cappella}, starts from high-level experience where participants mark states or events they want to understand and the system works toward finding patterns in physiological and other sensor data that correlate with those marks. \textbf{PatternFinder}, similar to~\citet{hartmann2007authoring}'s Exemplar system, starts from low-level signals where participants identify patterns in their physiological data that interest or puzzle them and translate them to discrete events.
However, whereas such systems have typically constrained their domain to examine the technical capabilities of such systems, we focus on how such added agency may or may not enable ontological boundary negotiation. 

To support open-endedness and flexibility, the probes use a Wizard of Oz approach to decouple the authoring experience from present-day technical constraints by instructing the participants to label their data for training a personalized model after the study is completed. 
This is because training real-time models on-device without constraining the domain was not possible at the time of this study and constraining the domain would have restricted boundary negotiation. 
However, a progress wheel was designed to communicate some indication of accumulated data for each category to the user through simulated algorithmic feedback, which acted as the wizard. 
In addition, the category creation screens were designed to remain flexible and revisable throughout the study, giving participants the ability to create, edit, and delete categories at any point.

\subsection{EventMarker}
\begin{figure}[h]
     \centering
     \includegraphics[width=\textwidth]{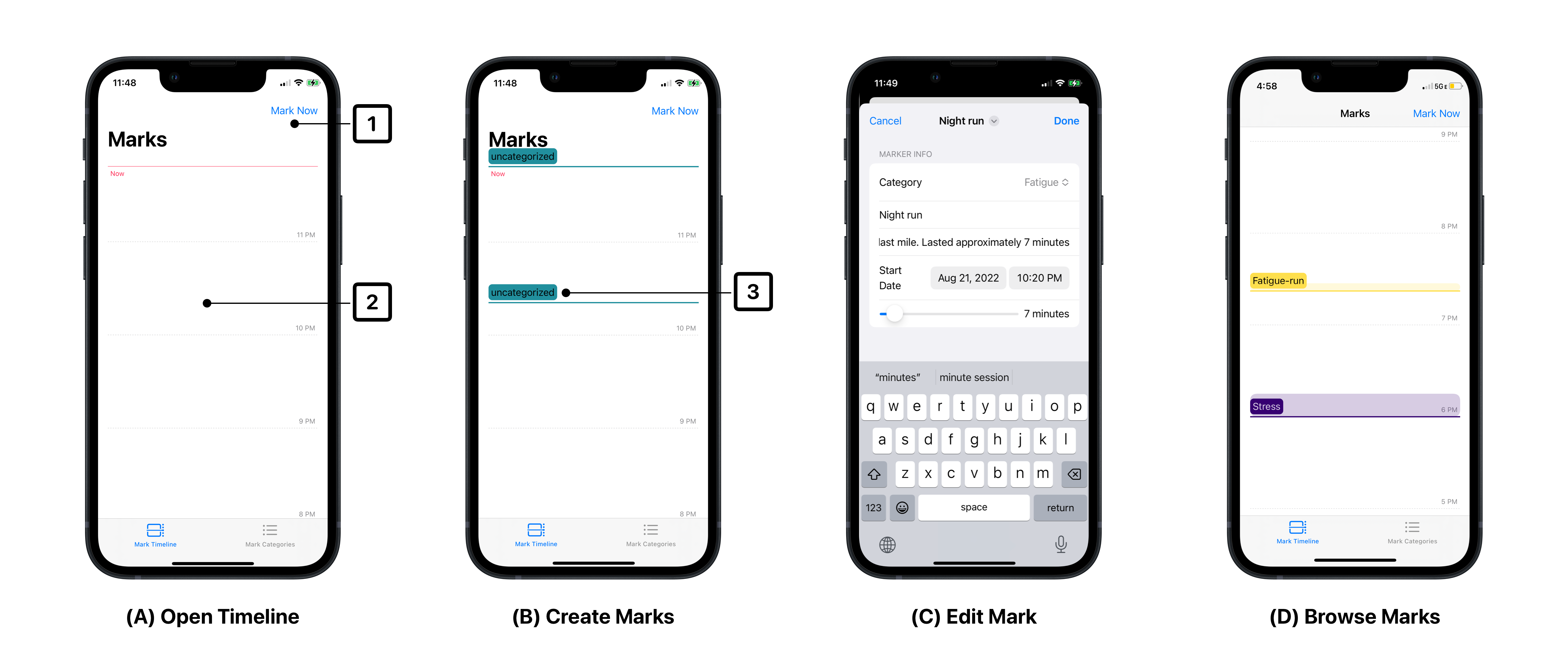}
     \Description[visual desc.]{A 4 phone screen user flow for the EventMarker app. The screens are labeled from left to right: (A) Open Timeline, (B) Create Marks, (C) Edit Mark, (D) Browse Marks). ``(A) Open Timeline'' is a screen of the app that's titled ``Marks.'' and shows an empty timeline that scrolls vertically, with the top marked ``Now'' and the bottom marked ``8 PM''. The current time shown is 11:48 PM. There is a button in the top right that says ``Mark Now''. There are two numbered annotations on this screen: (1), which points at the ``Mark Now'' button, and (2), which points at empty space in the middle of the timeline. It is explained in the figure caption that tapping either of these spaces will create a new mark, which leads to screen ``(B) Create Marks''. ``(B) Create Marks'' shows the same screen, but now with two ``uncategorized'' marks on it, displayed by two horizontal spanning dark green lines, one at ``Now'' and the other at about ``10:20 PM'' that are labeled ``uncategorized''. There is one annoation for this screen, (3), which points to one of the ``uncategorized'' marks. The caption for this figure states that tapping on this mark will lead you to screen ``(C) Edit Mark''. Screen ``(C) Edit Mark'' is a popup dialogue that prompts for mark info, including ``Category'' (filled in ``Fatigue'' here), a title (``Night run'' here), notes (``last mile. Lasted approximately 7 minutes'' here), a start date (``Aug 21, 2022: 10:20PM'' here) and a slider for the duration of the mark (``7 minutes'' here). Finally, screen ``(D) Browse Marks'' shows a the timeline screen with two categorized marks labeled ``Fatigue-run'' in yellow and ``Stress'' in purple. 
     ``(B) Create Marks'' }
     \caption{User flow for creating a mark in the EventMarker app. First, a user can browse the timeline by scrolling vertically, starting at ``now'' (A). They may use the ``Mark Now'' button (1) or tap anywhere on the timeline (2) to create a new uncategorized mark (B). In addition, a companion smart watch application enables access to the ``Mark Now'' button on the go. Next, they tap on a mark to edit it (3). On the subsequent modal (C), they may specify category, duration, and other information. They can also browse existing marks on the timeline (D).}
     \label{fig:flow_eventmarker}
 \end{figure}

EventMarker enables participants to mark states or events of interest in a timeline, which the system uses to find correlated patterns in their recorded physiological data. For example, a participant might mark every time they feel fatigued, and once training is complete, the system would surface patterns across their recorded data such as step count, sleep quality, or calendar information, associated with those marks. EventMarker participants were asked to create categories and mark states across any domain every time they take place: social relationships, environmental patterns, physical states like fatigue, or mental states like boredom. EventMarker consists of a mobile application and a companion smartwatch application to create marks without opening the phone application (see Figure~\ref{fig:flow_eventmarker}).

\subsection{PatternFinder}
\begin{figure}[h]
     \centering
     \includegraphics[width=\textwidth]{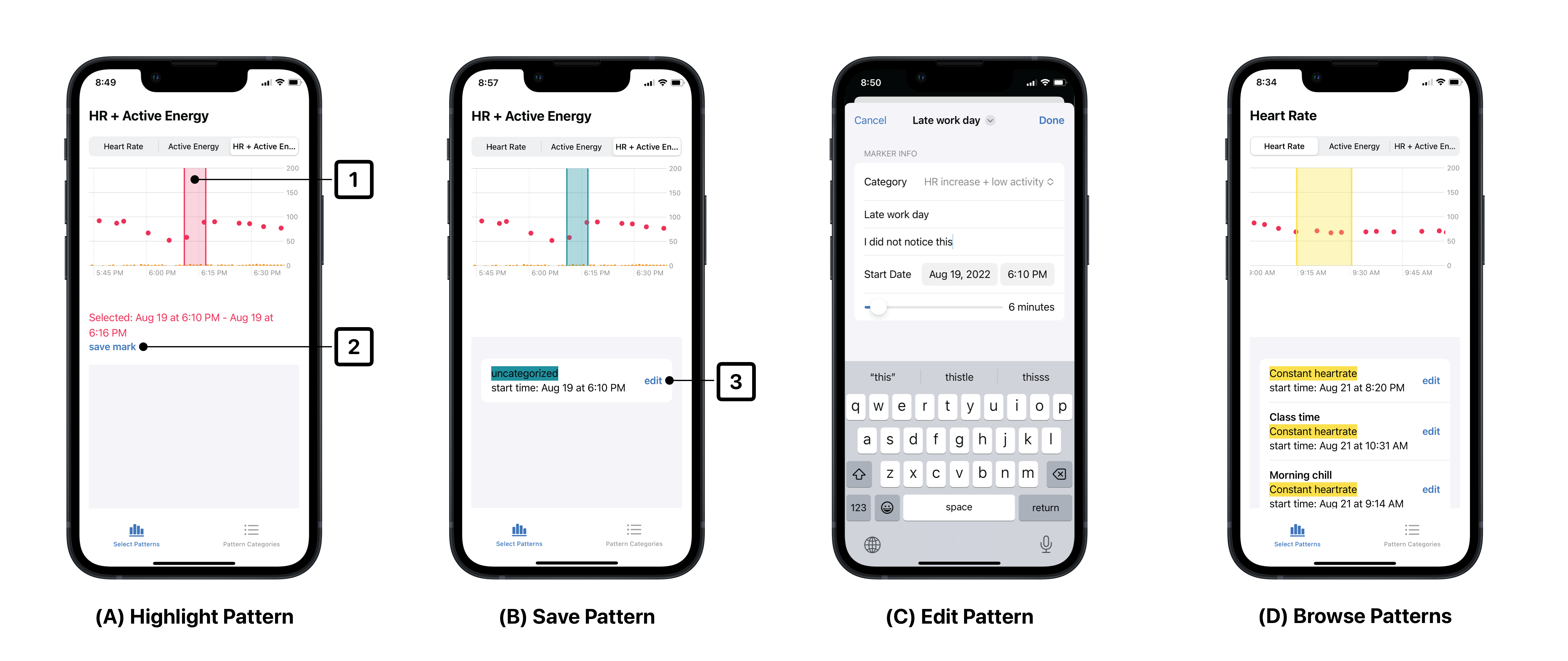}
     \Description[ivsual desc.]{A 4 phone screen user flow for the PatternFinder app. The screens are labeled from left to right: (A) Highlight Pattern, (B) Save Pattern, (C) Edit Pattern, and (D) Browse Patterns. The first screen, ``(A) Highlight Pattern'', is an interface titled ``HR + Active Energy'' at the top. Directly below this title is a tab-bar showing three options: ``Heart Rate'', ``Active Energy'', and ``HR + Active Energy'' (which is selected. Below this tab-bar is a scrollable chart with time on the X Axis and Heart Rates on the Y Axis. Plotted on this chart are pink circles representing heart rate readings, and small orange bars showing active energy at those times. There is a highlighted time range in the chart, shown as a vertically spanning transparent pink rectangular region. Below this chart is a section with a line of pink text that reads ``Selected: Aug 19 at 6:10 PM - Aug 19 at 6:16 PM" corresponding to the highlighted region. There is a button below this entry that reads ``save mark''. There are two annotations on this screen: (1), which points to the highlighted region, and (2) which points to the ``save mark'' button. The caption for this image shares that with (1) the highlighted region, a user taps and drags along a section of the timeline to select a pattern of interest, and they can save this pattern by tapping on (2) the save mark button. This will lead to screen ``(B) Save Pattern''. This screen is identical, except the previously pink highlighted region is now green, and instead of the "Selected: Aug 19..." text with the save mark button, there is an entry for an ``uncategorized'' mark with a ``start time'' of ``Aug 19 at 6:10PM'' and an edit button in that entry. This edit button is annotation (3), and the caption for this figure states that tapping this button will lead to screen ``(C) Edit Pattern''. Screen (C) shows a dialogue for editing pattern infromation, which includes ``Category'' (filled in with ``HR increase + low activity'' here), a title (filled in with ``Late work day'' here), a note (filled in with ``I did not notice this''), a Start Date (filled in ``Aug 19, 2022 6:10PM''), and a duration slider (set to ``6 minutes''). Finally, screen ``(D) Browse Patterns'' shows the timeline screen similar to screens (A) and (B), but with one yellow highlighted region (a pattern) and at least three entries below, all in yellow categorized as ``constant heartrate''.}
     \caption{User flow for selecting a pattern in the PatternFinder app.  First, the user can browse a horizontal timeline with their data plotted by scrolling (A) and once they find a pattern of interest, they tap and drag along a section of the timeline to select the pattern of interest (1). If satisfied, they save this period as a pattern (2), resulting in an uncategorized pattern appearing in a list (B). Next, they may tap to edit this pattern (3), specifying category, duration, and other information (C). A user can browse one of a few  timelines showing various data: ``Heart Rate'' (BPM), ``Active Energy'' (kcal), or heart rate overlaid over energy used (D).}
     \label{fig:flow_patternfinder}
 \end{figure}

PatternFinder enables participants to browse their physiological data, Heart Rate (HR), Active Energy, and HR + Active Energy overlaid, and select patterns that interest or puzzle them, which the system would use to learn to detect those patterns in real time. For example, a participant might create a category for instances where their heart rate is elevated but their activity is low, and over time, the system would detect and surface similar patterns. Participants were asked to browse their data at least once daily and to save patterns they wanted to understand better (see \figurename~{\ref{fig:flow_patternfinder}}).

\subsection{Shared Components}

Both applications enable the participants to create, delete, and edit events/patterns and categories. Furthermore, a progress wheel provided simulated feedback in place of actual feedback on model training progress. 

\subsubsection{Editing Events/Patterns}
On both applications, newly saved patterns are initially registered as \textit{uncategorized}. By tapping on the saved event or pattern, the user can add further annotations including Category (drop down menu consisting of the user’s predefined categories, for example, ``Fatigue (during a run)''), Descriptor/name (descriptor for that specific instance of the event, for example, ``morning mellow run''), Notes (to add additional information for themselves or to share with the researchers), Start date and time (to edit the start date and time using a slider/drop down), and Duration.
Once a category is assigned, the color will change to reflect the category color as defined by the user.

\subsubsection{Editing Categories}
\begin{figure}[h]
     \centering
     \includegraphics[width=\textwidth]{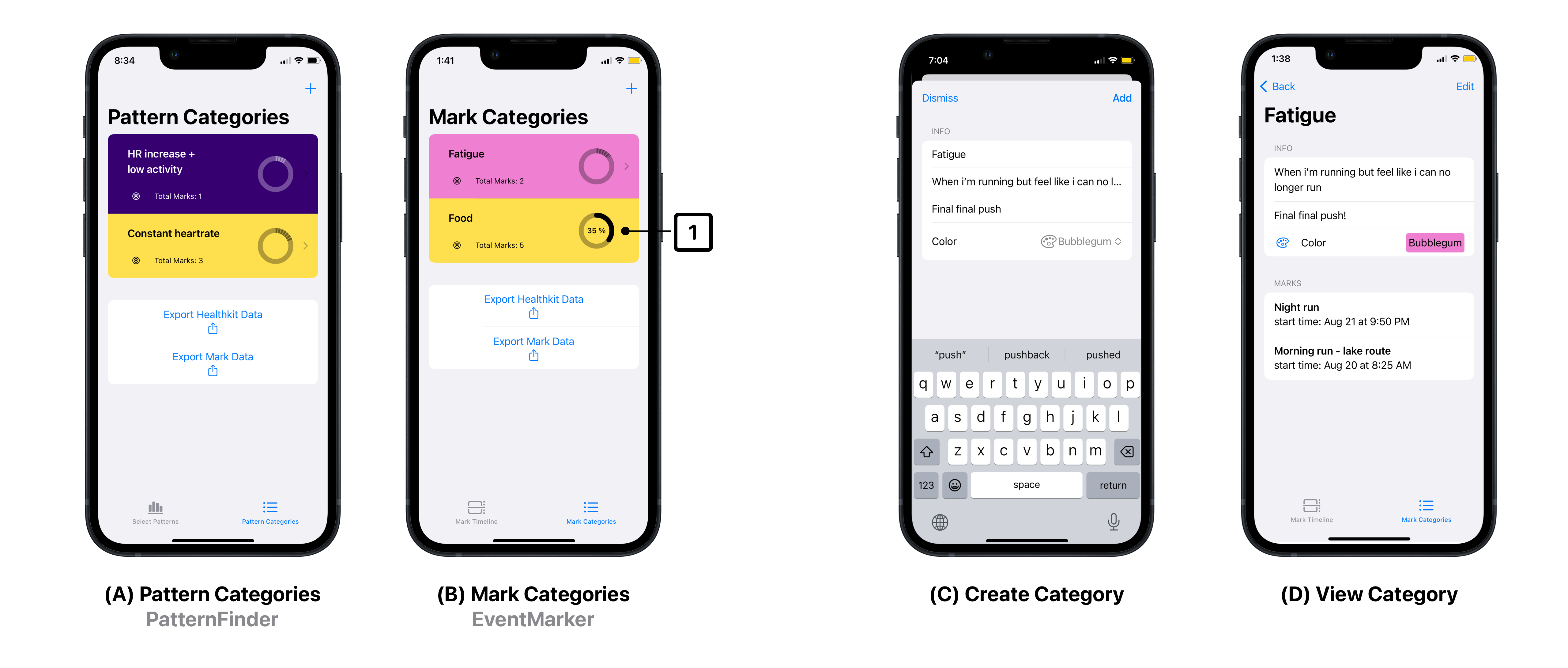}
     \Description[visual desc.]{A set of four images showing the phone screens for ``Category'' viewing and editing. From left to right, they are: (A) Pattern Categories (Pattern Finder), (B) Mark Categories (EventMarker), (C) Create Category, and (D) View Category. Screen ``(A) Pattern Categories''. The left most screen ``(A) Pattern Categories'' is a screen on in the PatternFinder app, accessible through the tab bar at the bottom, where ``Pattern Categories'' is currently selected, and ``Select Patterns'' is unselected. Below the title ``Pattern Categories'' there is a section of category entries, each with its own unique background color. The first one is ``HR increase + low activity'' and is themed purple. Below its name is text reading ``Total Marks: 1''. To the right is a progress wheel, with about a tenth of it filled in (representing model training progress for this  pattern). The next category is yellow, titled ``Constant heartrate'' and has similar elements. Below the categories are two buttons: ``Export Healthkit Data'' and ``Export Pattern Data''. The next screen, ``(B) Mark Categories'' is nearly identical, with two categories labeled ``Fatigue'' and ``Food'' in pink and yellow instead. The progress wheel for ``Food'' reads (35 percent). There is an annotation (1) next to this progress wheel, and the figure caption states that it denotes the completeness of a machine learning model learning that mark or pattern. Screen ``(C) Create Category'' shows a dialogue for entering information about a category, including a title (filled in ``Fatigue''), a description (filled in ``When I'm running but feel like I can no l...''), a note (filled in ``Final final push''), and a color (filled in ``Bubblegum'' from a drop down selection). Screen ``(D) View Category'' shows a screen for one of the categories (in this case ``Fatigue''). It includes an info section with the description, note, and color, as well as a section with a list of marks or patterns that have said category (in this case, 2 marks, one titled ``Night run'' with a time stamp, and another titled ``Morning run-lake route'' with a different time stamp.}
     \caption{In both PatternFinder and EventMarker, users create ``Categories'' for their patterns and marks. These are viewable from a  ``Categories'' screen: (A) and (B). Each category has a progress wheel indicating the completeness of a machine learning model learning that pattern or mark (1). Users may create and edit categories with custom names, descriptions, and colors (C), or view the information of an existing category, including all its marks or patterns (D).}
     \label{fig:flow_categories}
 \end{figure}

Both probes share a category screen where participants define and manage their own categories for classifying saved marks or patterns (see \figurename~{\ref{fig:flow_categories}}). Each category includes has four associated parameters of Category name (for example, ``Fatigue-run''), Info (for example, ``when I feel fatigued during a run''), Guide (notes for themselves to remember the logic, for example, ``only mark the final push before you can no longer run''), and Color (which gets reflected on all the data categorized under this category).

\subsubsection{Progress Wheel}
A progress wheel on each category provides simulated feedback on model training progress, designed to be intentionally ambiguous~\cite{gaver2003ambiguity} to prompt reflection on what progress might mean without anchoring participants to a specific interpretation. Every saved event or pattern gets assigned a random value of -5, 0, +5, or +10 (100 point scale) to mimic the behavior of a model being trained on-device. During pilot studies with two participants, the participants felt discouraged and confused when there were too many negative values assigned. Therefore, the probability of a positive value getting assigned was increased for the study. Additionally, the percentage count in the initial phases was removed and the display was changed to show the percentage numerically once the value has reached the 30\% threshold.

\subsubsection{Implementation}
Both applications were implemented as native iOS applications using Swift and SwiftUI. The applications integrated with Apple HealthKit to collect physiological data from participants' Apple Watches and iPhones. The visualizations were developed with SwiftCharts. All participant health and usage data was collected and stored locally on their phones, and transferred to a secure research laptop via AirDrop during the offboarding session.

\section{Exploratory Study}
The qualitative, exploratory nature of this study aligns with established traditions in design research that value situated insights over statistical generalizability~\cite{gaver2012should}. Thus, the aim was not to produce generalizable findings across populations but rather to explore a diverse range of potential sites for negotiation of existing ontological boundaries, and the factors that limited or contributed to such negotiation. 
% \subsection{Exploratory User Study}
Next, we describe the details of the week-long exploratory study with eight participants.
% Our intention was for the two open-ended design probes to scaffold the imagination of the participants, providing an opportunity to imagine possible uses beyond what today's technology feasibly allows. 

\subsection{Participants}
For recruitment, an initial screening survey was sent to a randomly selected pool of 600 participants at a large technology company. Selection criteria included adults 18 and older with basic data literacy (i.e., able to read and interpret a graph from a health application) who wear a smartwatch daily. The goal was to recruit a balanced pool across gender identity, age, and experience with personal sensing tools. Eight participants were recruited (3 female, 5 male), with an average age of 38 (range 28--47), varied professional backgrounds (HCI research: 1, engineering: 4, design: 1, IT: 1, project management: 1), and tracking technology usage (daily: 6, weekly: 1, monthly: 1). Although we intended to balance the participant pool across level of education completed, based on the people who filled out the screening survey, the lowest level of education completed for the participant pool was a bachelor's degree and six people had completed graduate education.
Participants received a \$40 gift card upon completion. 

\subsection{Procedures}
\begin{figure}[h]
     \centering
     \includegraphics[width=1\textwidth]{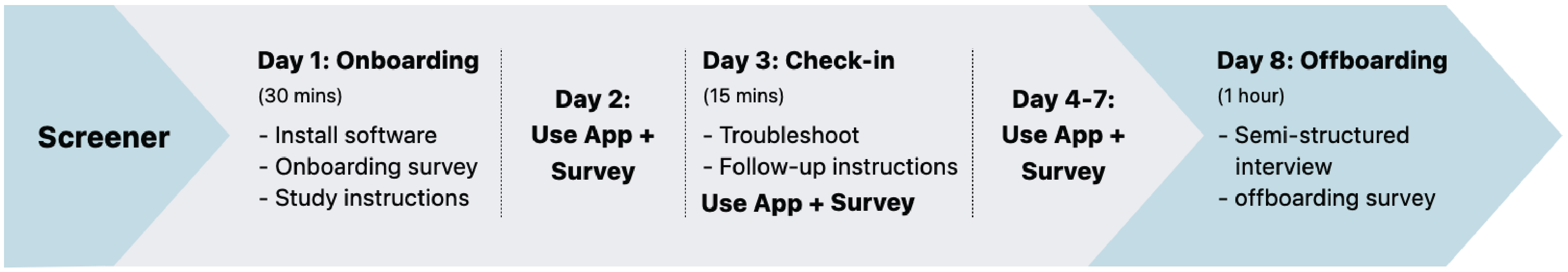}
     % \Description[visual desc.]{A timeline of the study procedure. On the left is the start, titled ``Screener''. Next is ``Day 1: Onboarding'', is noted to be 30 minutes, and includes installing software, an onboarding survey, and study instructions. Next is ``Day 2: Use App + Survey''. Next is ``Day 3: Check-in'' which is noted to be 15 minutes and includes troubleshooting and follow-up instructions. Day 3 also inclused ``Use App + Survey''. Next is ``Day 4-7: Use App + Survey''. Finally, ``Day 8: offboarding'', which is noted to be 1 hour and includes a semi-structured intreview and an offboarding survey.}
     \caption{Study procedures. The offboarding survey was completed prior to the semi-structured interview.}
     \label{fig:procedures}
 \end{figure}

The study ran for eight days and consisted of an onboarding session, a mid-study check-in on day three, daily surveys throughout, and an in-person offboarding session on day eight (see \figurename~{\ref{fig:procedures}}). The study was approved by the company's internal review board. 

Each participant was assigned to one of two probes rather than both to preserve the open-ended, exploratory character of the study. This decision was shaped by two factors: First, this setup allowed an examination of how the level of information abstraction shapes the boundaries people noticed or questioned. Second, because using both systems simultaneously risked shifting the participants' attention toward comparing the tools rather than engaging with phenomena and the act of boundary-drawing itself.

\subsubsection{Onboarding} Participants completed a pre-study survey that asked about their motivation for participation. A 30 minute onboarding session followed where they installed either application on their personal device(s), and were given instructions for using the system. 
Each participant received a copy of the study instructions.

EventMarker asked the participant to consider states or actions they want to find patterns in or gain a better understanding of, and leave a mark on the application every time those states or actions took place. Participants were encouraged to think about broader domains of life: \textit{``This can be in any domain that you want. It can be related to your social relationships and time you spend with certain people, environmental patterns like places you walk by, things that you do, like exercise or eating, physical states like fatigue, or mental states like boredom.''}

For PatternFinder, the following prompt was provided: \textit{``Every day, at least at the end of the day but more frequently as you wish, you look at your data and find patterns that are interesting to you and you would like to understand better.''}

\subsubsection{During the Study} Participants were instructed to use their assigned design probe for one week and complete a daily survey to report on any insights or problems they had encountered. On the third day of the study, a 15-minute check-in was conducted with each participant to address any technical or usability issues and answer any questions they had. 

\subsubsection{Offboarding} After seven days, each participant filled out an offboarding survey which included questions about usage patterns, usefulness of the system, usability of the system, and their general relationship to their data. Additionally, it included questions inspired by the Technology-Supported Reflection Inventory (TSRI) questionnaire \cite{tsri}. Given the exploratory and open-ended design of our systems, the questionnaire was modified to reflect how the user \textit{imagines} using the system. The answers to the survey were used to structure the semi-structured interview.
After the survey, the participants took part in an hour-long in-person offboarding session, where they transferred the study data and completed a semi-structured interview regarding their experience with the study and in-depth questions about the marks they had made and their imagined use. 

\subsubsection{Data Collection} In addition to the onboarding, offboarding, and daily surveys, the mark and category data from the applications as well as seven physiological signals including calories burnt during exercise, walking stability, walking asymmetry, heart rate, heart rate variability (HRV), VO2Max, and exercise time (anything more intense than a brisk walk) were collected. 

\subsubsection{Pilot Study} A pilot study with two participants was conducted with each using one of the two interventions for one day. Participants used the system and shared feedback on the two systems as well as the procedures. Using the insights from the pilot study, the systems were debugged, and the protocol was modified to improve the onboarding, add the day 3 check-in, and change the design of the progress wheel as described previously.

\subsection{Data Analysis}
We used the transcribed offboarding interviews and mid-study check-ins, daily survey responses, and mark, pattern, and category logs for data analysis. As a first step, we used affinity diagramming~\cite{beyer1999contextual, affinity} to group concepts across participants. Inductive open coding was used to examine the data and allow themes to emerge.
This initial pass gave an overview of how participants used the probes, the types of categories and use cases they imagined, and the challenges they encountered.

We then shifted to a diffractive reading of the data. 
As an analytical method, diffraction has its roots in feminist technoscience and the work of~\citet{minh2023not, minh1987difference}, later extended by~\citet{haraway2013promises} and~\citet{barad2007meeting, barad_diffracting_2014}.
In HCI, diffraction is increasingly used as a method~\cite{sanches2022diffraction, 10.1145/3706598.3713940}, and more recently, proposed as a method for ontological work~\cite{haghighi2025ontologies}.
We used a diffractive analysis method~\cite{barad2007meeting} because we aimed to identify moments of boundary negotiation rather than generalizable patterns of use. Because the researchers may also be naturalized to such boundaries, diffraction is a suitable method to bring to attention the patterns of difference to denaturalize what might be deemed natural, even for the researcher. 
For example, if two participants had the experience of abandoning a category, rather than reporting that under ``abandonment of category,'' a diffractive analysis focuses the attention on differences between the two experiences and the context in which they occurred, and what such difference may expose that would have otherwise remained hidden.
Reading diffractively through boundary negotiation revealed moments where a participant had to decide what does or does not belong to a category, whether a pattern was an ``interesting'' signal or noise, or where the tentativeness of boundaries undermined the usual authority of algorithmic detection. 
Importantly, tending to these patterns through diffraction turned experiences that may have normally been grouped under confusion, as a possibility for denaturalizing a normative boundary.

We emphasize that reading diffractively includes the researcher as an active participant in the analysis of the data. 
Thus, analyzing the data through the lens of ontological boundary negotiation, itself shapes and is shaped by what is considered an existing boundary, and what might be considered a negotiation of that boundary. 
Our research team included researchers and developers from industry and academia with backgrounds in design, neuroscience, machine learning, data visualization, sensing, and HCI, and we each bring our own assumptions and lived experiences to such analysis. The insights offered here are one set of possibilities, shaped by that positionality.

\section{Insights}
Next, we share the insights from the one-week exploratory study with the two open-ended probes, first by sharing narrative vignettes of participant experiences, followed by sharing the sites for boundary negotiation.
P1-P4 used EventMarker (EM) and P5-P8 used PatternFinder (PF).

\subsection{Participant Narrative Vignettes}
This section offers a brief account of the participant experiences in this study and the contexts that may have shaped their experience. Figures~\ref{fig:timelineall} give an overview of some of the participant logs.

\subsubsection{Motivation for Participation}
When asked about the reason for participating in the study in the onboarding session, P6 noted an interest in learning about themselves and P7 an interest in improving their current tracking routine. 
The others cited curiosity about the study or contributing to science. 
% We invite readers to attend to moments where participants encountered the edges of what they thought was trackable, meaningful, or possible, over time, and to the probe the factors that may have enabled or constrained those encounters.

\subsubsection{Participants that Frequently Use Personal Sensing Systems}
P3 (EM), an avid tracker, took the open-ended probe as an invitation to share their ideas around what was missing from their tracking experience. Throughout the study, they compared and contrasted EM with their existing personal sensing tools, and suggested UI and experience improvements (such as ways to make the experience ``sticky''). The categories they tracked followed from this background where they identified a possible but missed opportunity for personal sensing tools (automatically detecting a tic) and started to manually track it, imagining how it might soon get implemented in a product.
Despite their confidence in their existing practices, having to determine boundaries of ``exhaustion'' led them to notice subtle nuances in their experience they had not noticed before, for example, the differences and similarities between mental and physical exhaustion. 
They considered the progress wheel an attempt at gamification, but one that did not ``spark joy.''

P4 (EM) frequently interacts with their sensor data but only when training toward a goal, such as for a marathon. Using EventMarker, they enthusiastically shared that having to log ``events of interest'' without being bound to what they thought was ``trackable,'' an exciting experience. Over the course of the week, they found themselves being on the lookout for what an ``interesting'' pattern might be, and continued to mark things like eye strain or pulled muscle even though they did not think current sensors can detect them. 

P7 (PF) usually only looks at their data during fitness sessions, and has a dedicated fitness watch they use in addition to their daily smart watch. At first, they felt like the patterns they saw in their data ``matched what they already know.'' However on day four, they found instances of higher calorie burn but no corresponding increased heart rate, not knowing what triggers the calorie increase or how it may feel in their body. Additionally, they tracked a walk they go on with their dog every day that follows the same route to understand if the same walk reads the same in data or not. P7 also noted that their heart rate is much lower than average and they frequently get notified by their smart watch that their heart rate is ``dangerously low.''

P8 (PF) looks at their step count and weight daily, and occasionally monitors the steps climbed, however they engage with their data more passively compared to other frequent trackers. With PatternFinder, they initially had a hard time identifying ``interesting'' patterns, saying that it was ``hard to see what I'm looking for'' and initially found their data ``boring.'' At the time of the study, the Apple Watch did not regularly measure and log heart rate, which led P8 to mark those absences in their data as a pattern. They also felt like one week was too short ``By the time I get used to oh there is this pattern then the week is gone.'' 
However, their lack of confidence or experience ultimately led to them to experiment with the app.

\subsubsection{Participants that Occasionally Use Personal Sensing Systems }
P1 (EM) occasionally looks at their health data with their wife and some friends to see how their health is improving. In the study, they decided it will be ``too cluttered'' to track everything, so they chose to track events they are ``comfortable missing but are things you still want to do.''
During the study, they had an ``epiphany'' that they should be able to track events that are meaningful to them so they can decide what they want to do with them, even if that data will not make sense to developers or other people. They expected that by the end of the study the system should be able to automatically detect some of their marks and were disappointed that this was not the case. 
Interestingly, they noted that over time, they became ``More cognizant of how I specify my categories where I don’t need to recategorize them later. Before I hit save, does this actually categorize things I wanna do? Are these things co-related? Do I need to rebrand?''

P5 (PF), is not a regular tracker though they had tracked workouts in the past, and with the study they wanted to ``have fun.'' They took a systematic approach to exploring their physiological patterns. On the first day of the study, they went back and marked all of the patterns they found interesting in their prior week's data. This gave them the categories that they continued to use throughout the week, making no changes after the first exploration. P5 was judicial in their tracking during the study. Throughout the week, they developed an interest in new and relational understandings of their data; in relation to their partner, to an activity in terms of another activity, or to a physiological signal relating to a subjective perception of similar activities. 
They found the progress wheel motivating and expressed that they could have ``completed'' one of their categories if they had ``a little bit more time.''

% \begin{table}
% \renewcommand{\arraystretch}{1.5}
%   \begin{tabular}{ccl}
%     % \toprule
%     Design Probe&Category Groups&Category Names\\
%     \midrule
%     EventMarker & Activity — intended & search walk (P1), daily run (P1), chill walk (P1), playing outdoors (P1),  \\ & & passing time (P1), drink water (P3), food (P4)\\
%     % \cline{2-3}
%       & Event — unintended & binge eat (P2), rub nose (P3), scratchy throat (P3), pick / squeeze (P3)\\
%           % \cline{2-3}
%       & State / Feeling & tiring (P1), moods (P2), feel exhausted (P3), eye strain (P4), \\ & & pulled muscle (P4), need coffee (P4), fatigue (P4), muscle pain (P4)\\
%       \midrule
%     PatternFinder & Activity — intended& steps climbed (P8), yoga (P8), outdoor walk (P7), treadmill run (P7)\\    
%      & Pattern& medium intensity energy activity (P5), higher HR and moderate energy \\ & & (P5), sustained low HR (P5), up and down (P8), lower intensity energy \\ & & activity (P5), cluttered mid-range HR datapoints (P5), small peak in HR \\ & & quick recovery low energy (P5), high energy (P8), spike then decay (P6), \\ & & high cal no HR (P7), no activity (P8)\\
%         % \cline{2-3}
      
%   \bottomrule 
% \end{tabular}
% \caption{Category names recorded on the design probes after one week of use. The Category Groups column was identified by the research team based on the saved categories. 
% }
% \label{tab:use}
% \end{table}

\begin{table}
\small
\renewcommand{\arraystretch}{1.5}
\centering
\begin{tabularx}{\textwidth}{l l X}
% \toprule
\textbf{Design Probe} & \textbf{Category Groups} & \textbf{Category Names} \\
\midrule
EventMarker
  & Activity — intended & search walk (P1), daily run (P1), chill walk (P1), playing outdoors (P1), passing time (P1), drink water (P3), food (P4) \\
  & Event — unintended & binge eat (P2), rub nose (P3), scratchy throat (P3), pick/squeeze (P3) \\
  & State/Feeling & tiring (P1), moods (P2), feel exhausted (P3), eye strain (P4), pulled muscle (P4), need coffee (P4), fatigue (P4), muscle pain (P4) \\
\midrule
PatternFinder
  & Activity — intended & steps climbed (P8), yoga (P8), outdoor walk (P7), treadmill run (P7) \\
  & Pattern & medium intensity energy activity (P5), higher HR and moderate energy (P5), sustained low HR (P5), up and down (P8), lower intensity energy activity (P5), cluttered mid-range HR datapoints (P5), small peak in HR quick recovery low energy (P5), high energy (P8), spike then decay (P6), high cal no HR (P7), no activity (P8) \\
\bottomrule
\end{tabularx}
\caption{Category names recorded using the design probes after one week of use. The Category Groups column was identified by the research team based on the saved categories.}
\label{tab:use}
\end{table}

\subsubsection{Participants that Rarely Use Personal Sensing Systems}
P2 (EM) rarely tracks their data outside of the study. They intentionally do not wear their watch because it ``feels like too much technology connected.'' They only use a fitness application to record the number of days they workout. They got sick during the study, and noted that they were not able to speak very much during the offboarding session due to losing their voice. They had hoped that with the tool, they could see the reasons why they did something rather than simply tracking it. Having initially created a category called ``moods,'' they stopped using the category when they ``couldn’t put down a reason why I was feeling low.'' Whereas having marked binge eating, they realized that when they are accompanied by friends they are more likely to binge eat, an insight they found useful. Citing their sickness, P2 did not make any marks after the third day.

P6 (PF) wears an Apple Watch but has only used it to monitor their high heart rate in the past when they were ill. Initially, when they got their watch, they found the step count useful but they noted that they now know what contexts lead them to walk more so they no longer use this data. 
During the study, they ``never paid attention to active energy because it was always so low.'' They identified a single pattern in their heart rate (spike then decay) that they wish they could better understand in context to see what it would feel like. They also wished they could have more contextual data overlaid on the PatternFinder data (such as their calendar) to be able to make sense of what is signal and what might be noise.

\begin{figure}
     % \centering
     \includegraphics[width=1\textwidth]{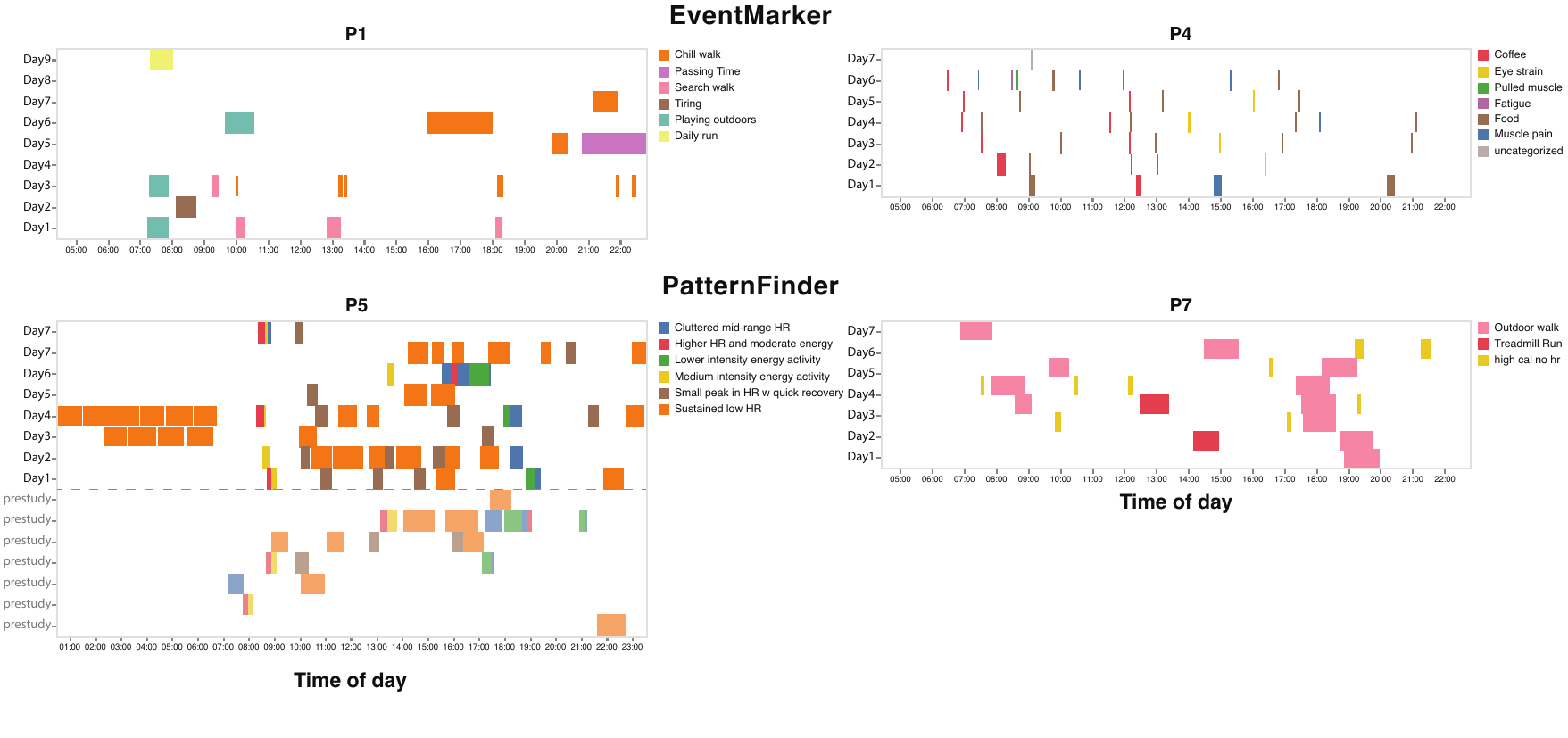}
     \Description[visual desc.]{A graphic showing usage pattern for participants. On top are two charts for EventMarker. Both have time of day on the x axis going from 5 AM until around 11PM and day of participant on the y axis. The top chart is titled ``P1''. P1 has 5 different categories of marks: chill walk, passing time, search walk, tiring, playing outdoors, and daily run. They have between 0 and 8 marks per day in the study, with 2 days of 5+ marks and 7 days of 4 or fewer marks. The most marked categories are ``Search Walk'', ``Chill Walk'' and ``Playing Outdoors''. P2 has 7 different categories of marks, and their marking consistency is more regular, with between 4 and 9 marks per day, except day 7 which has 1. There are many ``coffee'' and ``food'' marks throughout each day. On the bottom half of the graphic are two charts for PatternFinder. The first is labeled ``P5''. P5 has 6 categories of patterns through their 7 days in the study: cluttered mid-range HR, higher HR and moderate energy, lower intensity energy activity, medium intensity energy activity, small peak in HR w quick recovery, and sustained low HR. They also identified patterns from the 7 days leading up to the study. There are many patterns in each day, with the ``Sustained low HR'' being the most popular category. Pattern identification frequency per day looks bimodal, with about half the days having many patterns (10+) and half having few (5 or fewer). Below is P7, which has just 3 categories: outdoor walk, treadmill run, and high cal no hr. P5 highlighted 2 or fewer patterns on days 1, 2 and 7, and between 3 and 5 on days 3 through 6. Patterns are dispersed fairly uniformly through between 7 AM and 10 PM. }
     \caption{Usage pattern for participants from both groups. Each row displays the mark or pattern and the time of the day it was made. Note the different timescale for P5 since they wore their smart watch to sleep. 
     In EventMarker, participants had different strategies for making marks. P1 was consistent at adding duration whereas P4 stopped adding duration to their marks. In PatternFinder, note that P5 has a number of overlapping patterns. }
     \label{fig:timelineall}
     % \vspace{40.2244pt}
 \end{figure}

\begin{figure}[h]
     \centering
     \includegraphics[width=1\textwidth]{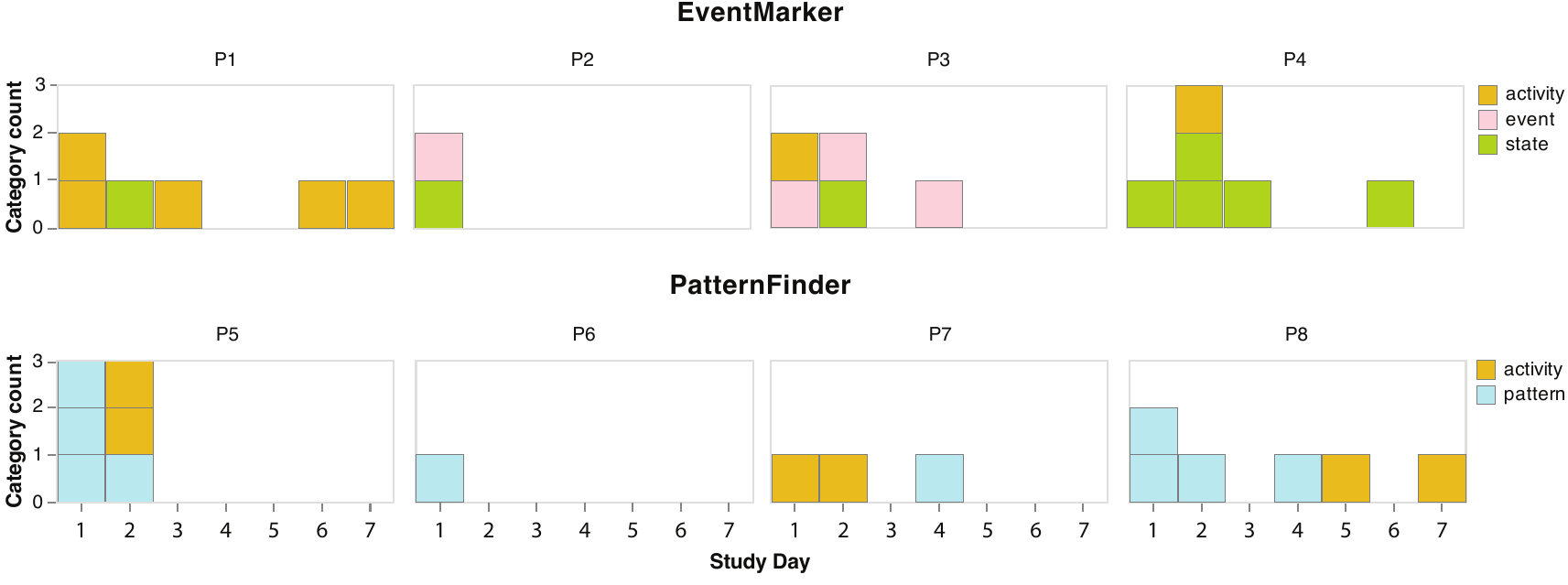}
     \Description[visual desc.]{A visualization of Mark and Pattern categories by participant. On top is a 2x2 grid of charts for EventMarker, with each chart labeled ``P1'', ``P2'', ``P3'', and ``P4'' in reading order from top left to bottom right. Time is shown on the X axis, from Day 1 to Day 7, and ``Category Count'' is shown on the Y Axis. P1 has 2 orange categories (activity) on day 1, one green one (state) on day 2, one activity on day 3, none on days 4 and 5, and one ``activity'' on days 6 and 7 each. P2 has just two categories on day 1, one green (state), and one pink (event). P3 has one activity and one event on day 1, one event and one state on day 2, nothing on day 3, one event on day 4, and nothing on days 5-7. P4 has one state on day 1, two states and an activity on day 2, one state on day 3, nothing on days 4 and 5, one state on day 6, and nothing on day 7. Below is another 2x2 grid of charts for PatternFinder, labeled P5 through P8. P5 has 3 ``patterns'' on day 1, 1 ``pattern'' and 2 ``activity'' on day 2, and nothing on days 3 through 7. P6 has one ``pattern'' on day 1 and nothing else. P7 has 1 acitivity on day 1, one activity on day 2, nothing on day 3, one pattern on day 4, and nothing on days 5 through 7. P8 has 2 patterns on day 1, 1 pattern on day 2, nothing on day 3, one pattern on day 4, one activity on day 5, nothing on day 6, and 1 pattern on day 7. }
     \caption{Mark and pattern category creation by participant. We thematically organized the created categories into four groups: ``Activity'' corresponding to intended activities such as a run or eating food, ``Event'' corresponding to unintended or unwanted events such as scratching the nose, ``State'' corresponding to a resulting state or feeling such as fatigue, and ``Pattern'' (only in PatternFinder) corresponding to an observed pattern in a sensor signal without connection to the contextual experience.\\ Note the evolution of categories. For example, P8 was first logging patterns but had a realization that marking activities can also be interesting. An important difference to note between the two interventions is that none of the participants in PatternFinder marked an event or state.}
     \label{fig:category_creation}
 \end{figure}

\subsection{Negotiating the Boundaries of a Phenomenon}
% The probes revealed that drawing a boundary is work, and that the work, when it became visible to participants, was where classification became living.

The probes invited the participants to define what they wanted to track, drawing a boundary between what the thing being tracked \textit{is} and \textit{isn't}. This led to instances of the boundary shifting and being negotiated over time. 
P1 (EM) began marking ``walks'' on their first day. Over the course of the week, this category made them realize that to them, a ``walk'' is actually two things; one is a ``chill walk'' where they know where they were going, and the other is a ``search walk'' where they are searching for an address. Thus, on the third day, they split the category into two by making a new ``chill walk'' category and renaming the original ``walking'' category to ``search walk'' (see Figure~\ref{fig:timelineall}). 

P3 encountered a similar experience with a category they called ``feeling exhausted,'' defining it as when they ``vaguely felt tired.'' In the offboarding interview, they noted that the sustained act of marking led them to recognize that their ``Feel exhausted'' category was flattening two distinct embodied states: ``I can be mentally exhausted but not physically, or physically but not mentally.'' 
Unlike P1, they did not split the category.

On the other hand, some phenomena did not lead to such negotiation or shifting of boundaries. For example, P3 also tracked their water intake but the boundary was drawn so neatly that over time, tracking water intake became irrelevant: ``as I was marking I found myself marking it a lot so it became redundant, almost like reporting on whether or not I took a deep breath.''  
Despite this, P3 continued to track their water intake, marking it a total of three times on the first, fifth, and seventh day. 

% The same dynamic played out at the level of the probes themselves. PatternFinder participants began from physiological signal where the data at times displayed edges and values and durations that left little room for negotiation. 
While EventMarker created the conditions for less clear boundaries to be drawn, PatternFinder reinforced the marking of more clear boundaries, leaving little room for negotiation. 
For example, P5 noted that when heart rate and active energy are overlaid, it leaves little room for negotiation: ``When you combine it’s clear that this is what I was doing.''
Unlike EventMarker, no PatternFinder participant created a category relating to a ``state/feeling,'' or ``event (unintentional)'' (see Table~\ref{tab:use} and Figure~\ref{fig:category_creation}). This could be because these categories have less well-defined boundaries, compared to something like an ``activity'' such as a run that has a clear start and end.

In EventMarker, participants noted how the act of marking events changed what they paid attention to. 
P3 noted that they became ``more aware of the habit because marking them, [makes me] recognize the habit or action more even when not marking it. I feel that I am noticing things more in my attention.'' P4 similarly found themselves ``thinking about their categories'' throughout the day. 
However, three of four EventMarker participants explicitly discussed how they restricted themselves to a limited number of categories despite being told they could track anything. 
P3 noted that their categories became ``front of mind.'' Therefore, they only created ``five to six'' categories because that felt ``doable.'' 

PatternFinder lent itself to a different type of interaction. P5, in a systematic first-day exploration of the prior week's data, established all the categories they would use and made no changes afterward: ``most recurring ones I was able to capture.'' The negotiation that drawing boundaries facilitated in EventMarker was largely not present in PatternFinder.
Beyond the affordance of the level of abstraction itself, the ``guides'' field, which were designed to help the participants maintain consistency of marks for ML model training purposes by taking notes, further pushed PatternFinder participants toward rigid and static boundaries. P5 developed exact rules such as ``heart rate between 100 and 150, medium intensity between 2 and 6'' and worked to apply them consistently: ``I was trying to be consistent, I'm sure I failed.''

The progress wheel may have produced a related effect. None of the participants reached 100\% in any of the categories. Three of the participants, felt negatively about not ``completing the circles'' as P1 called them. They felt ``sad'' to not be able to complete any of the circles and P7 felt like one week was ``too short'', and if they had more time that they could have ``made it to 100\% at least in one category.''
The progress wheel inadvertently gave each category a completable form, producing the effect that a boundary can be ``completed.'' 
For many participants, this completion resembled many ``gamification'' effects they were familiar with, making it a desirable state that may have foreclosed experimentation.

Negotiating the boundary of a phenomena may have been related to prior tracking experience. P8 who had a more passive experience with tracking, initially struggled to identify what counted as interesting: ``little hard to see what I'm really looking for. Maybe one week is too short. By the time I get used to oh there is this pattern then the week is gone.'' 
However, this lack of prior experience, led P8 to negotiate not just the boundaries of a given category but what the notion of ``interesting'' meant to them. 

Finally, technical constraints may have also contributed to boundary negotiation. P7's phone slowed down over the course of the study, which made it more difficult to modify categories because the interaction had become too slow.

\subsection{Negotiating the Subject (as Part of Relations)}
Although the study left the question of who the subject of tracking was open, we expected the subject to be the one typically reinforced by sensing systems: a bounded individual whose states and activities could be marked as properties of that individual alone. 
However, in some instances, participants negotiated the boundary of the subject being tracked. 
That negotiation pointed toward the relational circumstances that produced a state, the other beings a phenomenon was co-constituted with, and the temporal relations that only became visible across time.

In some instances, the idea that what was being tracked belonged to an individual was questioned altogether. 
P2 only cared to mark events alongside the relations they belonged to. 
When they marked binge eating, they found themselves marking it specifically when they were with a friend: ``most of the times when I'm with a friend I binge eat; when I'm alone it's a healthy choice.'' The phenomenon P2 was interested in was not binge eating as a property of P2 but binge eating as something that happened in a particular relational context. The social relations that brought the phenomenon into being were constitutive of the phenomenon being tracked. 
Similarly, they wanted to track low moods but what they wanted to track was not the mood itself but the circumstances that brought it into existence. 
When they could not identify those circumstances, they abandoned the category: ``Couldn't put down a reason why I was feeling low, so I stopped using that category.''

In other cases, the phenomenon being tracked did not belong to just the participant.
P7 walked their dog daily, following the same route. During the study they began marking these walks: ``walking with my pet, if route A or route B, sometimes we go clockwise and counterclockwise, if I'm beating my pace, if I'm consuming less or more energy. This week I consumed less energy because I was walking slower or it took me longer, might be the interesting part.'' Although they did not explicitly discuss this, the pace, the route, and the energy expenditure were shaped by the dog's behavior as much as P7's own. Rather than a property of P7, the walk fluctuated with both them and their dog together. 

Another relational negotiation of the object and subject boundary was where what was being tracked seemed to itself belong to an assemblage. 
P3 had been marking three tics (rub nose, scratchy throat, and pick/squeeze), initially considering each in isolation. Over the course of the week, they realized that the tics appeared together in temporal proximity, co-occurring in ways that brought into existence an assemblage P3 had little executive control over. 
Thus, P3 became interested in tracking how the shape of the three tics, shifted and evolved: ``If I could see that apps or the phone think they are correlated such that I could say I observe or don’t observe that and reflect on whether that’s sometime is missing or if I want to check on myself more.'' 
In this case, tracking the three tics was not just a matter of tracking the causal relationship between them, but in understanding the changes of a new and evolving phenomenon --- the tics as an assemblage --- that was taking shape and changing across time.
For example, at times, the different tics occurred in different combinations but in close proximity, other times by themselves, and P3 wondered about their correlation: ``but if it could know it, if there are correlations to something else to get that feedback, this other thing is on the rise.''

P3 pointed at similar assemblages across feelings and actions: ``interesting to see how actions or feelings and actions can correlate. For example I’m sad, then I go on a run, then I’m in peace. Or I feel good so I check in on my family and I feel good.''
The unit of interest in this case can be considered not as a state or an event that correlate neatly, but a sequence of feelings and actions across time that constitute a phenomenon \textit{together}. Thus the assemblage, not the individual elements, become the phenomenon.

\subsection{Negotiating Signal vs. Noise}
The probes made visible that the line between signal and noise is at times drawn even before participants see their data, and that some participants, in different ways, refused to accept where that line had been drawn. This negotiation was primarily seen in PatternFinder, where participants were directly interacting with physiological data as measured by sensors. 

P8 noticed that the Apple Watch did not regularly log heart rate. Rather than treating this as missing data, P8 began marking the absences themselves as a pattern: ``in the beginning I was like that's interesting, but then I'm thinking like hmm how do I know if this is real or not, so that kind of reduces the confidence in the pattern that I'm seeing. So there's that.''
However, in tracking the pattern, they realized that ``the heart rate, it doesn't monitor it when I'm sitting down.'' 
Because P8 did not have a lot experience interacting with their data, rather than dismissing absence as missing data, they turned it into a signal. In doing so, they ended up auditing the sensor's classification logic, examining the device's decision about what to record.

P7 marked the instances where active energy was recorded but where their heart rate did not increase which they called ``high cal no hr.'' Active energy and heart rate are both considered proxies for exertion, which P7, as an avid tracker was aware of. However, in seeing the two signals overlaid led to a breakdown of this expectation. Rather than treating it as an error or an anomaly, they considered the divergence of the two, a signal itself.

P6 identified only one pattern, marked by a spike followed by a rapid decay in their heart rate. They followed this pattern throughout the study: ``I just would want to understand, what does it mean? Is it meaningful in any way? And if it's telling me oh you're having one of these, I might be like, what am I doing right now? And I might be like oh here's what I'm feeling in this situation, and that might help me decide whether or not this is a pattern I care about.'' Rather than the system deciding what this signal meant, P6 wanted to decide for themselves whether it was noise or a signal worth attending to: ``what's noise, what's signal, what's typical, what's me.'' 
P6 also noted that their active energy was always so low that they had stopped paying attention to it in the study. Perhaps it was because of this that they had developed a mistrust for what the system may consider to be a signal or noise. 

Such disappearance however can be a result of falling neatly on \textit{both} sides of the boundary. 
P7 had a similar experience with heart rate where their chronically low heart rate triggered regular alerts from the Apple Watch that their heart rate was dangerously low: ``low heart rate, manually override or teach the device that that is normal for me so I don't wake up to the same alert.'' 
P7's inability to decide what is considered to be noise or signal for them eventually led to fatigue: ``there might be normal and giving that threshold some allowance end up creating alert fatigue and once you go into that it's hard to go back.'' When the boundaries of what gets measured become fixed, it risks becoming noise to those who fall outside those boundaries. When a boundary becomes noise, it either becomes easier to question \textit{or} it disappears altogether. 
Contrasting this, P7 described how the system's predictions with their own Garmin watch that they regularly use, eventually became so good that they ultimately stopped questioning it. Although that transition is supposed to be the goal, if the classification has no trace of its own history, it becomes more difficult to question, negotiate, or even perceive where it draws boundaries between signal and noise, regardless of which side of the boundary one fits. 

\subsection{Negotiating the Objectivity of Data}
A fourth site of boundary negotiation was around the objectivity of data, both in terms of the phenomena being detected or how such phenomena comes to be represented.  

The dominant approach to developing personal sensing systems is to build an algorithm for detecting a phenomenon that can be generalized and is largely legible to everyone using it such as calories burnt or number of steps~\cite{epstein2020mapping}. 
However, as capabilities evolve to enable people to build their own algorithms, there is no need for the phenomenon being detected to have a generalizable meaning for others. 
P1's insights around their own experience best articulates this: ``blackbox data I give the system that doesn't make sense to the system but makes sense to me.'' When classification systems are authored by others such as a system developer, they need to be legible to anyone using the system in the future. As people become able to define their own systems of classification, those systems need be meaningful only in relation to the person's own interpretive frame.
Importantly, this opens a broader possibility for creating classifications that do not necessarily fit into a bounded vocabulary even for the person creating them and instead, mark assemblages of feelings and events that may otherwise be inarticulable and see how and if they may algorithmically be detected and interpreted. 

Another interesting site for negotiating objectivity of data was related to how participants imagined encountering those boundaries.  
Although the probes did not train any algorithms on device, during the interview process, participants were asked how they imagined using a given algorithm once the training was completed. 
At times, participants wanted to understand their data not as absolute truths, but in relation to another person, to another activity, to another point in time, or to what was absent rather than present. 
In these cases, although the boundaries of the phenomena were not necessarily negotiated, how such boundaries were represented created possibilities for relationality. 
These relational translations of data offer opportunities for negotiating boundaries of phenomena by changing how they get interpreted, with meaning depending on what a boundary is put in relation to.

P8 imagined understanding activities in terms of each other, for example, understanding a swim in terms of weeks of daily walking: ``one thing that might be interesting is to be able to compare routine activities with once in a while activities. Not in terms of equivalence. Understand how much more daily activity compares to a longer activity. Helps you actually have a better estimate of how much you're stretching beyond daily routine.'' 
For P7, they were interested in understanding their outdoor runs in relation to their treadmill runs.

A related imagination was an attempt to understand differences between similar activities. 
P8 marked identical yoga sessions with the intention of detecting anomalies in a routine the system would otherwise register as identical each time, understanding each session not as an absolute measure but in relation to all the previous ones.

Other forms of relational understandings of data also emerged. 
P5 was interested in understanding their data in relation to their partner's: ``comparing calories we both burn and looking at diverging and aligned patterns.'' The interesting unit was not P5's calorie expenditure as an absolute value but the relationship between P5's patterns and their partner's when both were doing the same activity, with meaning existing in the relation between them.
Given that such relations could shift, a relational translation of data can offer a possibility for making classification less authoritative, foregrounding the interpretive work that can at times be invisible. Furthermore, such translation builds new types of evolving relationships between physical bodies that can change how we relate to other people (and can even be extended to how we relate to non-humans).
At the same time, this relational understanding may ultimately forget its own politics as well.
P7 discussed calibrating patterns between people to develop a common language. But calibration is not neutral. Thus, it is important to consider who and what becomes the point of reference, and who and what is always being compared to, or otherwise, how can centering one's experience be avoided through design.
\section{Discussion}

The results of the week-long study with eight participants using the two open-ended probes read diffractively, offered examples of negotiation of ontological boundaries and insights into what may have enabled such negotiations. In this section, we discuss what it means to create the conditions for boundary negotiation in and through design in personal sensing, discuss the politics of boundary negotiation, and close by discussing this work as a method for ontological design.

\subsection{Supporting Ontological Boundary Negotiation in Personal Sensing}
% negotiated when wasnt an activity, more hazy. how can we support attending to the hzy phenomena
The four negotiations --- of the boundary of the phenomenon, the subject, signal vs. noise, and the objectivity of data --- demonstrate that when given the possibility, people can start to notice and negotiate the ontological boundaries imposed on them by existing systems. 
However, external factors such as prior experience and technical limitations, as well as design decisions imposed by the system itself, may shape what negotiations become possible. 
Next, we share possible sites where design can support a negotiation of ontological boundaries in open-ended sensing systems.
This work extends prior work around flexibility and customization in personal sensing, both in defining what is being explored~\cite{karkar_tummytrials_2017} and what gets displayed~\cite{barker-canler_flexible_2024, PIphenomenology}, to offer ways in which flexibility and customization can support ontological negotiation. 

\subsubsection{Supporting Negotiation of how Phenomena are Defined}
Design can support the negotiation of boundaries of phenomena. Whereas normally, phenomena are considered to be objectively existing and measurable~\cite{leahu2014freaky}, design can communicate the tentativeness of these boundaries by supporting actions that allow a given boundary to evolve and change over time. 
One such support can be at the level of phenomena itself, supporting merging, splitting, or connecting categories to one another, or building and rearranging hierarchies into categories. 
Another area would be around what gets considered the subject of tracking. Rather than assuming a singular or objective subject, design can bring attention to how the phenomena being tracked can be inherently relational, as in the walk with P7 and their dog. 

Finally, artifacts can embed the tentativeness of boundary drawing in their design.~\citet{star1999sorting} call for classification to be ``living'' where users can see the politics of their construction by keeping the boundary-making decisions intact: ``A key for the future is to produce flexible classifications whose users are aware of their political and organizational dimensions and which explicitly retain traces of their construction'' (p. 326). 
To this end, demonstrating how boundaries already evolve and shift can better communicate this provisionality and enable questioning.
% Design can support boundary negotiation : i.e., merging, splitting, hierarchy of categories, expanding the “subject” beyond the individual, carry traces and histories of boundary-making decisions.

Beyond interface-level choices, the level of information abstraction itself can shape this perception of provisionality. EventMarker was more likely to bring participants' attention to the moment, to the boundaries of their experience, and to what resisted being named, whereas PatternFinder brought their attention to certainty and objective truth. To this end, no PatternFinder participant created a state, feeling, or event category; phenomena with fuzzier boundaries than a well-defined activity. Thus, the design and the level of abstraction at which participants encounter phenomena, shapes what can get negotiated.

\subsubsection{Algoithmic Support of Boundary Negotiation}
Algorithms hold authority over how phenomena get defined. As in the progress wheel, how the algorithm itself works shaped whether or not a boundary stayed fixed or got negotiated. 
The design of the progress wheel gave the understanding that each category is contributing to ``completing'' one algorithm that can ultimately detect what is being marked. However, even if the boundaries of phenomena are cleanly defined, which they are not, an algorithm would still be bringing some aspects of reality into view while ignoring others. To this end, algorithms and how their functions are communicated visually, can be designed to support boundary negotiation. For example, supporting forking an existing algorithm to experiment with how it can be modified or remixed, supporting comparison of different algorithms, and defining progress as more than reaching a neat end-state can be starting points for algorithmic support of boundary negotiation. 

% On the other hand, as generative AI is given large amounts of data to make sense of, it is interesting to facilitate 

% Algorithmic support of “Living Classifications”: i.e., forking, comparing, facilitating progress beyond focusing on an “end-state” to allow evolution of categories.

\subsubsection{Supporting Shifting Boundaries through Relational Meaning Making}
The other side of a predefined reality being objectively measured is the reality being objectively represented. A relational representation of data can interrupt this perceived objectivity, making room for negotiating boundaries. 

The participant accounts offered ample examples for moving away from this objective representation, to understanding phenomena --- and through that, ourselves --- relationally. 
For example, P8 expressed a desire to understand their ``once in a while activity'' in terms of the more regular ones. Taking this one step further, one possibility would be to use a similar mechanism to find sensor-based equivalents to different phenomena. For example, if a model is trained to detect when one is meditating, and the same model predicts with high certainty that the person is meditating when they are focused at work or in a heated conversation, it can create the possibility for reflecting on the similarities of the two states. 
Then, what would it mean to represent one activity in terms of other? What would that mean for people's understanding of themselves, as well as the understanding of the inner-workings of the algorithm and the politics of personal sensing more broadly, based on what activities come to be considered as similar?

P1's insight around blackbox categories offers a different possibility. While for P8, the system could connect one describable activity to another, for P1, blackbox data --- data that ``doesn't make sense to the system but makes sense to me'' --- does not have to be describable or correspond to a known linguistic idea. Instead, the blackbox categories offer a unique possibility for experimenting with states that have no language to describe them. 
% Furthermore, P1 had an insight around blackbox 
% drawing on P1's insight around blackbox categories that have no corresponding discrete experiences that can be described but are , 

However, with relational meaning making, we caution that there is also an increased risk of harm.
In their discussion on the politics of classification,~\citet{star1999sorting} remind us that ``whose story has categorical ascendancy here is forever morally moot. All of the stories are important and all of the categories tell a different one.'' 
Relational representations of the self may allow new stories to be told, but it also risks centering one story as more important. 
Thus, the goal and challenge becomes how to center multiplicities of stories, and to treat all as equally important and true. 
Furthermore, representing one state in terms of another can carry a risk of people consider one of those states as negative or undesirable. For example, if an undesirable state such as fear has a corresponding physical state to a desirable state such as loving one's partner, this can create anxiety, in particular if the person deems boundaries as objective and fixed.

\subsection{The Politics of Boundary Negotiation}
Although creating the conditions for boundary negotiation can facilitate ontological expansion, it can also inadvertently lead to harm.

In this study, at times, participants sought fixed boundaries.
P8 articulated that they want to ``establish [patterns] and forget about them.'' Fixed boundaries reduce ongoing cognitive work. They allow the body to recede from conscious attention --- what~\citet{leder1990absent} calls the dysappearing body --- which itself brings value. Boundary negotiation keeps the body present as an object of inquiry, and this presence is differentially burdensome across different people and different relationships to their bodies. 
Tracking the same yoga flow over time for P8 can lead to interesting insights around the other environmental or personal factors that may contribute to physiological change. 
However, if their physiology remains mostly the same, a level of sudden change may lead to concerns that may or may not be justifiable: ``I'm not sure about the real-time aspect of it. It can become scary. But it can be useful to have a weekly or monthly trend.'' 
P8's personal experience with a health condition that ``accumulated over time'' and became ``life threatening'' illustrates how ontological expansion around health data carries both potential benefits and risks. P8 valued the possibility of earlier detection through pattern recognition, but recognized how constant monitoring might generate anxiety or false alarms.

In addition, the labor of boundary negotiation is itself not equally distributed~\cite{power_li}. Participants with prior and more active tracking experience navigated the open-ended exploration more easily than those with less experience. Similarly technical challenges, such as P7's application slowing down over time, affected boundary negotiation. Thus, it is crucial for design to support and offer the infrastructure needed for boundary negotiation, support ongoing revisions, and in particular, scaffold and center the experiences of those already marginalized~\cite{karasti_infrastructuring_2014}.

At the same time, it is important to recognize that boundary negotiation is not always a \textit{choice} or a \textit{privilege}. For those who are in-between, or who fall on the margins of different boundaries, negotiating boundaries to make oneself legible is not a choice. 
What would it mean for a system to leave a trace of the boundary-drawing decisions, as a resource for navigating the ongoing negotiations some bodies unwantedly get enrolled in.
What would it mean to design systems that hold classifications alive~\cite{star1999sorting} and distribute the labor of keeping them alive with care. 
How can existing boundaries be communicated as provisional while maintaining the sense of safety and security that allows the body to dysappear when needed~\cite{leder1990absent}. 
Rather than a permanent invitation to keep all boundaries open, this work offers a path for making boundaries negotiable, and to communicate their provisionality.

\subsection{Open-ended Probe Study as a Method for Ontological Design}

In this work, participants at times negotiated boundaries, and in other cases, dominant sensing approaches shaped their imagination. Although this work surfaced the ways in which design can further support boundary negotiation, this method also presents a promising direction for ontological design. 

Critical and speculative design frequently takes a taken-for-granted boundary and brings it to one's awareness to be questioned~\cite{speculativedunne,purpura2011fit4life, odd_interpreters}. 
However, a challenge with this approach, and with ontological design more broadly, is that ontological boundaries are so naturalized that it is at times difficult to see or notice them, and imagine what they might be otherwise. To this end, methods such as breakdowns --- the ``interrupted moment of our habitual, standard, comfortable `being-in-the-world'''~\cite[p.77]{winograd1986understanding} --- have been used to surface a boundary previously naturalized~\cite{sabie2022unmaking, haghighi2023workshop}. 
Even then, imagining what else a boundary might be can be a challenge.

The method described in this paper presents a different path by starting from the dominant approaches without even needing to articulate exactly what a boundary might be. By making existing boundaries available for negotiation through open-ended systems over time, people may start to negotiate those boundaries. Conducting a diffractive analysis to identify moments of boundary negotiation as such then offers glimpses --- albeit at times small --- at what other boundaries might look like. 
To be clear, the negotiations we identified were likely not even perceived by the participants as a negotiation of existing boundaries. To a different research team, they may or may not read as negotiation either. However, through reading the results and looking for moments of boundary negotiation, no matter how small, we can identify those moments as openings. These openings can then be used in two ways; further expanded by designers, or supported through design to be further negotiated by people. 

For example, P3's insight with the three tics as being related led to our team noticing an emerging relational entity that did not have a predictable shape but contained shared elements occurring over time. Although P3 may have never articulated their experience as discovering an emerging relational entity, in us seeing it as such we are able to consider how design can support that imagination. For example, what might an algorithm designed to detect the assemblage with imperfect parts look like? What temporal properties or data resolution may contribute to seeing the patterns of such an assemblage in ways that current approaches to algorithm development may not? 

Similarly, the user interface can be designed to support P3 perceiving their tics not as isolated phenomena but as relational assemblages. For example, the application can allow the participants to mark ``related'' events; events that may feel related to the person but not have a measurable similarity. Or, as in the case of P2 and their friend who was a part of the binge eating event, contextual information such as other people involved in the event can be marked. 
Taking this approach has the potential for making this method cyclical. Each cycle can uncover boundaries previously unimagined, supporting such shifting through design, and shifting the boundaries further through a next round of people using the systems. 

\subsection{Design After Design: Ontological Openings when AI is Doing the Designing}
As prior work has discussed~\cite{10.1145/3706598.3713940, haghighi2025ontologies}, AI systems already \textit{do} prioritize some decisions and responses over others. Thus, the structure for the study described in this paper aimed to resemble the type of decisions that will likely also be embedded in sensing systems built with AI, even if the person imagined a personal sensing system with an AI, ``from scratch.''
As generative AI and malleable software~\cite{10.1145/3772318.3790713,litt2025malleable} increasingly put the act of designing systems in users' hands, one hope is that the ontological assumptions those systems carry will begin to reflect users' own commitments rather than those of system developers. 

This work offers a methodological approach for examining and attuning to the ontological openings that authoring agency makes possible, and attending carefully to what these openings reveal.
At the same time, it remains an open question how design should support ontological expansion in cases where AI is doing much of the designing. By implicitly embedding many decisions, such systems may shape the users' imagination even down the line. 
% Looking further ahead, as AI systems move toward generating sensing tools on demand, one might ask whether abstracting the core ontological considerations that get identified, 
One path forward is to identify principles, for example from this study, where the boundaries of phenomena gets drawn, who the subject is, what counts as signal, and what the data is allowed to mean, so that the tools these AI-based systems produce carry those ontological considerations forward rather than reproducing existing paradigms by default.  

\section{Conclusion}
This paper examines how design can create and support the conditions for ontological boundary negotiation.
Through two open-ended probes and a week-long study with eight participants, we shared empirical insights into four sites for boundary-negotiation in user-authored personal sensing systems.
Participants negotiated the boundaries of phenomena, the subject as a part of relations, what counted as signal and what gets considered as noise, and the objectivity of data. At times, these moments were partial and temporary, and tended toward fixity. 
We further discussed how design contributed to and may further facilitate and support such boundary negotiations.  

Importantly, the insights demonstrated that more agency does not automatically lead to boundary negotiation. 
Thus, as AI makes user authorship of sensing systems increasingly accessible, ontological assumptions will not diversify simply because users are doing the designing. Expanding beyond existing boundaries requires tools and practices for seeing where the openings are, attending to what they reveal about the boundaries presumed by the systems, and considering how design might support expanding beyond those boundaries.
To this end, this approach contributes a method for ontological design by creating the conditions for expanding beyond existing boundaries through open-ended engagement with technologies that would draw and fix boundaries otherwise.

% \begin{acks}
% \end{acks}

\bibliographystyle{ACM-Reference-Format}
\bibliography{references}

\appendix

\end{document}